\documentclass[aps,preprint,amsmath,amssymb,showpacs,amsfonts]{revtex4}
\usepackage{epsfig}
\usepackage{graphicx}
\usepackage{dcolumn}
\usepackage{bm}

\begin{document}

\title{Penrose Inequality for Asymptotically AdS Spaces}

\author{Igor Itkin}

\author{Yaron Oz}

\affiliation{Raymond and Beverly Sackler School of Physics and Astronomy, Tel-Aviv
University, Tel-Aviv 69978, Israel}

\date{\today}
\begin{abstract}
In general relativity, the Penrose inequality relates the mass and
the entropy associated with a gravitational background. If the inequality
is violated by an initial Cauchy data, it suggests a creation of a naked
singularity, thus providing means to consider the cosmic censorship
hypothesis. We propose a general form of Penrose inequality for asymptotically
locally $AdS$ spaces.
\end{abstract}

\pacs{04.20-q, 04.70.-s, 11.25.Tq }
\maketitle


\section{Introduction}

The Penrose inequality \cite{Penrose:1973um} (for reviews see e.g.
\cite{Bray:2003ns,Mars:2009cj}) relates the mass and the horizon
area of Cauchy initial data that if violated suggests  a creation of a
naked singularity.
Consider, for instance, the evolution in time of an asymptotically
flat initial Cauchy data with a mass $M_i$ and horizon area $A_i$.
The solution to Einstein equations with this initial data is expected
to settle down at a late time to a Kerr black hole with mass
$M$ and horizon area $A$. The Kerr black hole satisfies the
inequality (we use $c=G_{N}=1$) $M\geq \sqrt{A/{6 \pi}}$, which is saturated by the Schwarzshcild black
hole. The event horizon area does not decrease with time $A\geq A_i$
\cite{Hawking:1971tu}, while the mass cannot increase and may only
decrease due to radiation loss $M\leq M_i$. Thus,
\begin{equation}
M_i\geq M\geq\sqrt{A/{16\pi}}\geq\sqrt{A_i/{16\pi}}\ ,
\end{equation}
 and the initial Cauchy data should also satisfy the Penrose inequality
$M_i\geq\sqrt{A_i/{16\pi}}$. The argument relies on the Hawking
area theorem and the relaxation at late times to a Kerr solution,
both assume the weak censorship hypothesis (for a review see \cite{Wald:1997wa}). Therefore, a Cauchy data
that violates the Penrose inequality is likely to generate a naked
singularity.

In view of the AdS/CFT correspondence (for a review see \cite{Aharony:1999ti})
and the interest in higher-dimensional black holes, it is of interest to have a general form of Penrose
inequality for asymptotically $AdS$ spaces.
A special form of such inequality has been considered recently in the context of the fluid/gravity correspondence in \cite{Oz:2010wz}.
The purpose of this note is to construct a general form of the Penrose inequality for
asymptotically locally $AdS_{d+1}$ spaces, with
electrical charge and a
general boundary topology.

Consider the Einstein equations
\begin{equation}
G_{AB}+\Lambda g_{AB}+8\pi T_{AB}=0,~~~~~ A,B=0,...,d \ , \label{eq:Einstein_charged_d-1}
\end{equation}
where $G_{AB}$ is the Einstein tensor, $\Lambda =  -\frac{d(d-1)}{2 l^2}$ is a negative cosmological constant and
$T_{AB}$ is the stress-energy tensor for the electromagnetic field $F^{AB}$, whose field equation is $\nabla_{A}F^{AB}=0$.
The equilibrium background that saturates the inequality is the electrically charged black brane solution with charge $q$. Its background takes the form
\begin{equation}
ds^{2}=-fdt^{2}+f^{-1}dr^{2}+r^{2}d\Omega_{k,d-1}^{2},~~~~~~ F^{tr}=\frac{q}{r^{d-1}} \ ,
\end{equation}
where
\begin{equation}
f=k+\frac{r^{2}}{l^{2}}-\frac{m}{r^{d-2}}+\left(\frac{q}{r^{d-2}}\right)^{2} \ ,
\end{equation}
and $d\Omega_{k,d-1}^{2}$ is the $(d-1)$-dimensional metric on a flat space, a sphere or
hyperboloid for $k=0$, $k=1$ and $k=-1$ respectively.
We denote the corresponding $(d-1)$-dimensional volume by $\Omega_{k,d-1}$.

In order to derive the Penrose inequality we have to construct the entropy $S$ and mass $M$ associated with the background.
The entropy density $s$ reads
\begin{equation}
s=\frac{R^{d-1}}{4} \ , \label{entropy}
\end{equation}
where $R$ is the horizon location, $f(r=R)=0$.

The energy density $\varepsilon$ can be obtained from the $T_{00}$ component of the $d$-dimensional boundary stress-energy tensor \cite{Balasubramanian:1999re}.
One writes the $(d+1)$-dimensional metric in an ADM-like decomposition
\begin{equation}
ds^2 = N^2dr^2 + \gamma_{\mu\nu}(d x^{\mu} + N^{\mu}dr)(d x^{\nu} + N^{\nu}dr) \ ,
\end{equation}
where $\gamma_{\nu \mu}$ is the boundary ($r=const.$) metric
\begin{equation}
\gamma_{\mu\nu}d x^{\mu} d x^{\nu} = -f d t^2 + r^{2}d\Omega_{k,d-1}^{2} \ ,
\end{equation}
and $N$ and $N^\mu$ are the lapse and shift functions.
The boundary stress-energy tensor takes the form
\begin{equation}
T_{\mu\nu}=\frac{l}{8\pi}\lim_{r\rightarrow\infty}r^{d-2}\left(\theta_{\mu\nu}-\theta\gamma_{\mu\nu}\right) + counter~~terms \ ,
\label{T}
\end{equation}
where $\theta_{\mu\nu}$ is the extrinsic curvature tensor of the boundary
\begin{equation}
\theta_{\mu\nu}=-\frac{1}{2}\sqrt{f}\partial_{r}\gamma_{\mu\nu} \ ,
\end{equation}
and
$\theta=\gamma^{\mu\nu}\theta_{\mu\nu}$.

One gets
\begin{equation}
\varepsilon=\frac{(d-1)}{16\pi}\left[\frac{q^{2}}{R^{d-2}}+kR^{d-2}+\frac{R^{d}}{l^{2}}\right]+\varepsilon_{0} \ ,
\label{energy}
\end{equation}
where $\varepsilon_{0}$ is the Casimir energy \cite{Emparan:1999pm}
\begin{equation}
\varepsilon_{0}=\frac{1}{8\pi}(-k)^{d/2}\frac{(d-1)!!^{2}}{d!}l^{d-2}
\end{equation}
for $d$ even and zero for $d$ odd.

Using (\ref{entropy}) and (\ref{energy}) we get the
equilibrium relation between the energy density, the entropy density and electric charge:
\begin{equation}
\varepsilon=\varepsilon_{0}+\frac{d-1}{16\pi}\left[q^{2}\left(\frac{1}{4s}\right)^{\frac{d-2}{d-1}}+k\left(4s\right)^{\frac{d-2}{d-1}}+\frac{1}{l^{2}}\left(4s\right)^{\frac{d}
{d-1}}\right] \ .
\end{equation}

Going away from equilibrium we obtain the inequality relation among the mass $M$, area $A$ and charge $q$
\begin{equation}M - M_{0} \geq \frac{(d-1)\Omega_{d-1,k}}{16\pi}\left[q^{2}\left(\frac{\Omega_{d-1,k}}{A}\right)^{\frac{d-2}{d-1}}
+k\left(\frac{A}{\Omega_{d-1,k}}\right)^{\frac{d-2}{d-1}}+\frac{1}{l^{2}}\left(\frac{A}{\Omega_{d-1,k}}\right)^{\frac{d}{d-1}}\right] \label{gf}
\end{equation}
where
\begin{equation}
M_{0}=\frac{1}{8\pi}(-k)^{d/2}\frac{(d-1)!!^{2}}{d!}l^{d-2}\Omega_{d-1,k}  \label{ca}
\end{equation}
for $d$ even and zero for $d$ odd.
The form (\ref{gf}) together with (\ref{ca}) is our proposal
for a general form of the Penrose inequality for
asymptotically locally $AdS_{d+1}$ spaces, with
electric charge $q$ and a
boundary topology characterized by $k$.
The inequality (\ref{gf}) has been formally established when $d=3$ by Gibbons, using the inverse mean curvature flow \cite{Gibbons:1998zr}.
Partial forms of (\ref{gf}) were written in \cite{Gibbons:2005vp,Gibbons:2006ij}, where evidence for its correctness was given.

\section*{Acknowledgements}

We would like to thank I. Bakas and G. Gibbons for e-mail correspondence.
The work is supported in part by the Israeli Science Foundation center
of excellence, by the US-Israel Binational Science Foundation (BSF),
and by the German-Israeli Foundation (GIF).

\end{document}